\RequirePackage{ifpdf}
\ifpdf 
\documentclass[pdftex]{sigma}
\else
\documentclass{sigma}
\fi

\begin{document}

\allowdisplaybreaks

\renewcommand{\PaperNumber}{002}

\FirstPageHeading

\ShortArticleName{Singularity Analysis and Integrability of a Burgers-Type System of Foursov}

\ArticleName{Singularity Analysis and Integrability\\ of a Burgers-Type System of Foursov}

\Author{Sergei SAKOVICH~$^{\dag\ddag}$}

\AuthorNameForHeading{S. Sakovich}

\Address{$^\dag$~Institute of Physics, National Academy of Sciences, 220072 Minsk, Belarus}
\EmailD{\href{mailto:saks@tut.by}{saks@tut.by}}

\Address{$^\ddag$~Max Planck Institute for Mathematics, Vivatsgasse 7, 53111 Bonn, Germany}

\ArticleDates{Received October 28, 2010, in f\/inal form December 24, 2010;  Published online January 04, 2011}

\Abstract{We apply the Painlev\'{e} test for integrability of partial dif\/ferential equations to a~system of two coupled Burgers-type equations found by Foursov, which was recently shown by Sergyeyev to possess inf\/initely many commuting local generalized symmetries without any recursion operator. The Painlev\'{e} analysis easily detects that this is a typical {\it C}-integrable system in the Calogero sense and rediscovers its linearizing transformation.}


\Classification{35K55; 37K10}

\section{Introduction}\label{s1}

The system of two coupled Burgers-type equations
\begin{gather}
w_t   = w_{xx} + 8 w w_x + ( 2 - 4 \alpha ) z z_x , \nonumber\\
z_t   = ( 1 - 2 \alpha ) z_{xx} - 4 \alpha z w_x + ( 4 - 8 \alpha ) w z_x - ( 4 + 8 \alpha ) w^2 z - ( 2 - 4 \alpha ) z^3 ,\label{e1}
\end{gather}
where $\alpha$ is a parameter, was discovered by Foursov \cite{Fou} as a nonlinear system which possesses generalized symmetries of orders three through at least eight but apparently has no recursion operator for a generic value of $\alpha$. Foursov \cite{Fou} noted that two systems equivalent to the cases $\alpha = 0$ and $\alpha = 1$ of \eqref{e1} had already appeared in \cite{Svi} and \cite{OS}, respectively, and found a recursion operator for the system \eqref{e1} with $\alpha = 1/2$. Very recently, Sergyeyev \cite{Ser} proved that the system \eqref{e1} does possess an inf\/inite commutative algebra of local generalized symmetries but the existence of a recursion operator~-- of a reasonably ``standard'' form~-- for a generic value of $\alpha$ is disallowed by the structure of symmetries. Sergyeyev \cite{Ser} found that the algebra of generalized symmetries of \eqref{e1} is generated by a nonlocal two-term recursion relation rather than a recursion operator.

In the present paper, we explore what the Painlev\'{e} test for integrability, in its formulation for partial dif\/ferential equations \cite{WTC,Tab,RGB}, can tell about the integrability of this unusual system \eqref{e1} with $\alpha \neq 1/2$, which possesses inf\/initely many higher symmetries without any recursion operator. The Painlev\'{e} test easily detects that this is a typical {\it C}-integrable system, in the terminology of Calogero \cite{Cal}. In Section~\ref{s2}, we show that the singularity analysis of the Burgers-type system \eqref{e1} naturally suggests to introduce the new dependent variable $s(x,t)$,
\begin{gather} \label{e2}
s = z^2 ,
\end{gather}
to improve the dominant behavior of solutions. The system \eqref{e1} in the variables $w$ and $s$ passes the Painlev\'{e} test for integrability successfully: positions of resonances are integer in all branches, and there are no nontrivial compatibility conditions at the resonances. In Section~\ref{s3}, we show that the truncation of singular expansions straightforwardly produces the transformation
\begin{gather} \label{e3}
w = \frac{\phi_x}{4 \phi} , \qquad s = \frac{a^2}{( 4 - 8 \alpha ) \phi}
\end{gather}
to the new dependent variables $\phi(x,t)$ and $a(x,t)$ satisfying the triangular linear system
\begin{gather} \label{e4}
a_t = ( 1 - 2 \alpha ) a_{xx} , \qquad \phi_t = \phi_{xx} + a^2 .
\end{gather}
This linearizing transformation was found in an inverse form in \cite{TW} and used in a form close to~\eqref{e2},~\eqref{e3} in \cite{Ser}. Section~\ref{s4} contains concluding remarks.

\section{Singularity analysis}\label{s2}

First of all, let us note that the cases $\alpha = 1/2$ and $\alpha \neq 1/2$ of the system \eqref{e1} are essentially dif\/ferent, at least because the total order of the system's equations is dif\/ferent in these cases, and the general solution of this two-dimensional system contains dif\/ferent numbers of arbitrary functions of one variable in these cases, three and four, respectively. When $\alpha = 1/2$, the system~\eqref{e1} is the triangular system
\begin{gather} \label{e5}
w_t = w_{xx} + 8 w w_x , \qquad z_t = - 2 z w_x - 8 w^2 z ,
\end{gather}
where the f\/irst equation is the linearizable Burgers equation possessing the Painlev\'{e} proper\-ty \cite{WTC}, whereas the second equation simply def\/ines a function  $z(x,t)$ by the relation $z = f(x) \exp \int{\left( - 2 w_x - 8 w^2 \right) d t}$, with $f(x)$ being arbitrary, for any solution $w(x,t)$ of the Burgers equation. Thus, integrability of this case is obvious.

In the generic case of the Burgers-type system \eqref{e1} with $\alpha \neq 1/2$, we substitute into \eqref{e1} the expansions
\begin{gather} \label{e6}
w = w_0 (t) \phi^\sigma + \cdots + w_r (t) \phi^{\sigma + r} + \cdots , \qquad z = z_0 (t) \phi^\tau + \cdots + z_r (t) \phi^{\tau + r} + \cdots,
\end{gather}
where $\phi_x (x,t) = 1$, in order to determine the dominant behavior of solutions near a movable non-characteristic manifold $\phi (x,t) = 0$ and the corresponding positions of resonances. In this way, we obtain the following four branches, omitting the ones corresponding to the Taylor expansions governed by the Cauchy--Kovalevskaya theorem:
\begin{gather}
\sigma = \tau = -1 , \qquad w_0 = \frac{1}{2} , \qquad z_0 = \pm \sqrt{\frac{1}{4 \alpha - 2}} , \qquad r = -2, -1, 1, 2 ; \label{e7} \\
\sigma = \tau = -1 , \qquad w_0 = 1 , \qquad z_0 = \pm \sqrt{\frac{3}{2 \alpha - 1}} , \qquad r = -4, -3, -1, 2 ; \label{e8} \\
\sigma = -1 , \qquad \tau = - \frac{1}{2} , \qquad w_0 = \frac{1}{4} , \qquad \forall\, z_0 (t) , \qquad r = -1, 0, 1, 2 ; \label{e9} \\
\sigma = -1 , \qquad \tau = \frac{1}{2} , \qquad w_0 = \frac{1}{4} , \qquad \forall \, z_0 (t) , \qquad r = -1, -1, 0, 2 . \label{e10}
\end{gather}

We see that the system \eqref{e1} does not possess the Painlev\'{e} property because of the non-integer values of $\tau$ in the branches \eqref{e9} and \eqref{e10}. Nevertheless, the positions of resonances are integer in all branches, and we can improve the dominant behavior of solutions by a simple power-type transformation of the dependent variable $z$, just as we did for the Golubchik--Sokolov system in~\cite{Sak}. We introduce the new dependent variable $s$ given by \eqref{e2}, and this brings the Burgers-type system \eqref{e1} into the form
\begin{gather}
w_t   = w_{xx} + 8 w w_x + ( 1 - 2 \alpha ) s_x ,\nonumber \\
s s_t   = ( 1 - 2 \alpha ) s s_{xx} - \frac{1}{2} ( 1 - 2 \alpha ) s_x^2 - 8 \alpha s^2 w_x + ( 4 - 8 \alpha ) w s s_x \nonumber\\
\phantom{s s_t   =}{} - ( 8 + 16 \alpha ) w^2 s^2 - ( 4 - 8 \alpha ) s^3 .\label{e11}
\end{gather}
This form is hardly simpler than the original one, but the studied system \eqref{e1} in this form \eqref{e11} will pass the Painlev\'{e} test.

We substitute into \eqref{e11} the expansions
\begin{gather} \label{e12}
w = w_0 (t) \phi^\sigma + \cdots + w_r (t) \phi^{\sigma + r} + \cdots , \qquad s = s_0 (t) \phi^\rho + \cdots + s_r (t) \phi^{\rho + r} + \cdots ,
\end{gather}
with $\phi_x (x,t) = 1$, and f\/ind the following four branches:
\begin{gather}
\sigma = -1, \qquad \rho = -2 , \qquad w_0 = \frac{1}{2} , \qquad s_0 = \frac{1}{4 \alpha - 2} , \qquad r = -2, -1, 1, 2 ; \label{e13} \\
\sigma = -1 , \qquad \rho = -2 , \qquad w_0 = 1 , \qquad s_0 = \frac{3}{2 \alpha - 1} , \qquad r = -4, -3, -1, 2 ; \label{e14} \\
\sigma = \rho = -1 , \qquad w_0 = \frac{1}{4} , \qquad \forall \, s_0 (t) , \qquad r = -1, 0, 1, 2 ; \label{e15} \\
\sigma = -1 , \qquad \rho = 1 , \qquad w_0 = \frac{1}{4} , \qquad \forall \, s_0 (t) , \qquad r = -1, -1, 0, 2 . \label{e16}
\end{gather}
Now the exponents of the dominant behavior of solutions, as well as the positions of resonances, are integer in all branches.

The next step of the Painlev\'{e} analysis is to derive from \eqref{e11} and \eqref{e12} the recursion relations for the coef\/f\/icients $w_n$ and $s_n$ ($n = 0, 1, 2, \dotsc$) and then to check the compatibility conditions arising at the resonances. Omitting tedious computational details of this, we give here only the result. The compatibility conditions turn out to be satisf\/ied identically at the resonances of all branches \eqref{e13}--\eqref{e16}, hence there is no need to introduce logarithmic terms into the expansions~\eqref{e12} representing solutions of the system \eqref{e11}. The function $\psi (t)$ in $\phi = x + \psi (t)$ remains arbitrary in all branches. Also the following functions remain arbitrary: $s_1 (t)$, and either $s_2 (t)$ if $\alpha = 1$ or $w_2 (t)$ if $\alpha \neq 1$, in the branch \eqref{e13}; either $s_2 (t)$ if $\alpha = 3/2$ or $w_2 (t)$ if $\alpha \neq 3/2$, in the branch~\eqref{e14}; $s_0 (t)$, $s_1 (t)$ and $w_2 (t)$ in the branch \eqref{e15}; and $s_0 (t)$ and $w_2 (t)$ in the branch \eqref{e16}. The generic branch is \eqref{e15}: the expansions \eqref{e12} contain four arbitrary functions of one variable in this case, thus representing the general solution of the system~\eqref{e11}.

Consequently, the Burgers-type system \eqref{e1} in its equivalent form \eqref{e11} has passed the Painlev\'{e} test for integrability.

\section{Truncation technique}\label{s3}

There is a strong empirical evidence that any nonlinear dif\/ferential equation which passed the Painlev\'{e} test must be integrable. The test itself, however, does not tell whether the equation is {\it C}-integrable (solvable by quadratures or exactly linearizable) or {\it S}-integrable (solvable by an inverse scattering transform technique). Often some additional information on integrability of the studied equation, such as its linearizing transformation, Lax pair, B\"{a}cklund transformation, etc., can be obtained by truncation of the Laurent-type expansion representing the equation's general solution \cite{WTC,Wei,SKS,MC,KS,KKSST,CM,Hon}.

Let us apply the truncation technique to the system \eqref{e11}. We make the truncation in the generic branch \eqref{e15} which corresponds to the general solution. In what follows, the simplifying reduction $\phi = x + \psi (t)$, $w_n = w_n (t)$ and $s_n = s_n (t)$ ($n = 0, 1, \dotsc$) is not used. We substitute the truncated expansions
\begin{gather} \label{e17}
w = \frac{w_0 (x,t)}{\phi (x,t)} + w_1 (x,t), \qquad s = \frac{s_0 (x,t)}{\phi (x,t)} + s_1 (x,t)
\end{gather}
to the coupled equations \eqref{e11}, equate to zero the sums of terms with equal degrees of $\phi$, and in this way obtain the def\/initions
\begin{gather} \label{e18}
w_0 = \frac{\phi_x}{4} , \qquad s_0 = \frac{\phi_t - \phi_{xx} - 8 w_1 \phi_x}{4 - 8 \alpha}
\end{gather}
for the coef\/f\/icients $w_0$ and $s_0$, as well as a system of four nonlinear partial dif\/ferential equations for three functions, $w_1 (x,t)$, $s_1 (x,t)$ and $\phi (x,t)$. Two of the four equations of that system are the same initial equations \eqref{e11} with $w$ and $s$ replaced by $w_1$ and $s_1$, respectively, which means that the obtained system and the relations \eqref{e17} and \eqref{e18} constitute a so-called Painlev\'{e}--B\"{a}cklund transformation relating a solution $( w_1 , s_1 )$ of \eqref{e11} with a solution $( w , s )$ of \eqref{e11}. The other two equations of the obtained system are fourth-order polynomial partial dif\/ferential equations~-- let us simply denote them as $E_1$ and $E_2$ because it is easy to obtain them by computer algebra tools but not so easy to put them onto a printed page~-- they involve the functions $w_1$, $s_1$ and $\phi$ and contain, respectively, 55 and 110 terms.

Fortunately, there is no need to study the obtained complicated system of four nonlinear equations for compatibility in its full form. Instead, let us see what will happen if we take $w_1 = 0$ and $s_1 = 0$, which means that we apply the obtained Painlev\'{e}--B\"{a}cklund transformation to the trivial zero solution of the system \eqref{e11}. The reason to do so consists in the following empirically observed dif\/ference between {\it C}-integrable equations and {\it S}-integrable equations, which, as far as we know, has never been formulated explicitly in the literature. A Painlev\'{e}--B\"{a}cklund transformation of a {\it C}-integrable equation, being applied to a single trivial solution of the equation, produces the whole general solution of the equation at once. Examples of this are the Burgers equation \cite{WTC} and the Liouville equation in its polynomial form $u u_{xy} = u_x u_y + u^3$ \cite{SKS}. On the contrary, numerous examples in the literature show that a Painlev\'{e}--B\"{a}cklund transformation of an {\it S}-integrable equation, being applied to a single trivial solution of the equation, produces only a class of special solutions of the equation, usually a rational solution or a one-soliton solution with some arbitrary parameters (see, e.g., \cite{KS,KKSST,Hon,EP}).

Taking $w_1 = 0$ and $s_1 = 0$, we f\/ind that the equation we denoted as $E_1$ is satisf\/ied identically, whereas the equation we denoted as $E_2$ is reduced to
\begin{gather} \label{e19}
( \phi_t - \phi_{xx} ) ( \phi_t - \phi_{xx} )_t + \frac{1}{2} ( 1 - 2 \alpha ) ( \phi_t - \phi_{xx} )_x^2 - ( 1 - 2 \alpha ) ( \phi_t - \phi_{xx} ) ( \phi_t - \phi_{xx} )_{xx} = 0 .
\end{gather}
The general solution of this fourth-order equation contains four arbitrary functions of one variable, which is exactly the degree of arbitrariness of the general solution of the system \eqref{e11}. For this reason, we conclude that the system \eqref{e11} must be {\it C}-integrable. Now it only remains to notice that, if we introduce the new dependent variable $a (x,t)$ such that
\begin{gather} \label{e20}
\phi_t - \phi_{xx} = a^2 ,
\end{gather}
the equation \eqref{e19} becomes linear:
\begin{gather} \label{e21}
a_t = ( 1 - 2 \alpha ) a_{xx} .
\end{gather}
Finally, combining the relations \eqref{e17}, \eqref{e18}, \eqref{e20}, \eqref{e21} and $w_1 = s_1 = 0$, we obtain the exact linearization \eqref{e3} and \eqref{e4} for the system \eqref{e11}.

\section{\label{s4}Conclusion}

In the present paper, we used the Painlev\'{e} test for integrability of partial dif\/ferential equations to study the integrability of a system of two coupled Burgers-type equations discovered by Foursov, which possesses an unusual algebra of generalized symmetries as was shown by Sergyeyev. The Painlev\'{e} analysis easily detected that the studied Burgers-type system is a typical {\it C}-integrable system in the Calogero sense and rediscovered its linearizing transformation. As a~byproduct, we obtained a new example conf\/irming the empirically observed dif\/ference between  the Painlev\'{e}--B\"{a}cklund transformations of {\it C}-integrable equations and {\it S}-integrable equations. In our opinion, the Painlev\'{e} test deserves to be used more widely to search for new integrable nonlinear equations, because with its help one can discover new equations possessing such new properties which look unusual from the point of view of other integrability tests.

\subsection*{Acknowledgements}

This work was partially supported by the BRFFR grant $\Phi$10-117. The author also thanks the Max Planck Institute for Mathematics for hospitality and support.

\pdfbookmark[1]{References}{ref}
\LastPageEnding

\end{document}